\documentclass[namedreferences,hyperref,optionalrh]{spr-sola}
\usepackage{graphicx}        
\usepackage{color}           
\usepackage{float}

\chardef\us=`\_

\begin{document}

\begin{frontmatter}
\title{Height Variations of Magnetoacoustic Cutoff Frequency in the Solar Atmosphere}
\author[addressref={aff1},corref,email={virat.com@gmail.com}]{\inits{Pradeep}\fnm{Pradeep}~\snm{Kayshap}\orcid{0000-0002-0786-7307}}

\author[addressref=aff1,email={gayathrihegde05@gmail.com}]{\inits{Gayathri}\fnm{Gayathri}~\snm{Hegde}}

\author[addressref=aff2,email={}]{\inits{Z. E.}\fnm{Z. E.}~\snm{Musielak}\orcid{987-654-3210}}

\author[addressref=aff3,email={}]{\inits{Kris}\fnm{K.}~\snm{Murawski}}

\author[addressref={aff4,aff5},email={}]{\inits{Tobias}\fnm{Tobías}~\snm{Felipe}}

\address[id=aff1]{School of Advanced Sciences and Languages, VIT Bhopal University, Kothrikalan, Sehore, Madhya Pradesh - 46611}

\address[id=aff2]{Department of Physics, University of Texas at Arlington, Arlington, TX 76019, USA}

\address[id=aff3]{Institute of Physics,
            University of M. Curie-Sk{\l}odowska, 
             Pl.\ M.\ Curie-Sk{\l}odowskiej 1, 
             PL-20-031 Lublin, Poland}

\address[id=aff4]{Instituto de Astrofísica de Canarias, 38205 La Laguna, Tenerife, Spain}

\address[id=aff5]{Departamento de Astrofísica, Universidad de La Laguna, 38205, La Laguna, Tenerife, Spain}

\runningauthor{Kayshap et al.}
\runningtitle{\textit{Cutoff frequency with height}}
\begin{abstract}
The determination of the cutoff frequency in real solar observations under different local physical conditions is an important and insufficiently explored aspect of waves in solar physics. This work utilizes the near ultraviolet (NUV) spectrum of the QS, observed by the Interface Region Imaging Spectrograph (IRIS) on  November 16$^{th}$, 2013, in sit-n-stare mode. It contains several absorption and emission lines that form at different heights between the photosphere and chromosphere. Cross-wavelet analysis is performed on Doppler velocity time series of pairs of spectral lines sampling different atmospheric layers to estimate the cutoff frequency at six different heights between the photosphere and chromosphere. It is found that the cutoff frequency increases with height from around 3.0 mHz at 0.38 Mm (photosphere) to around 8.5 mHz at 1.2 Mm (chromosphere). Higher chromospheric heights show indications of standing oscillations. The presented observational results are compared with those previously obtained, and serve as a benchmark to refine theoretical models that predict variations of cutoff frequencies in the solar atmosphere. 
\end{abstract}
\end{frontmatter}
\section{Introduction} \label{sect:intro} 
The solar atmosphere reveals a broad spectrum of different waves, which are primarily generated in the solar convection zone, and after propagation through the solar photosphere, they may transfer their energy to outer atmospheric layers (e.g., \citealt{1979ApJ...231..570L, 1982ApJ...253..367L, 2011ApJ...727...17F, 2006ApJ...640.1153C, 2016A&A...585A.110K, 2023LRSP...20....1J}). This wave energy transfer is strongly affected by the presence of local cutoff periods in the solar atmosphere as these cutoffs may prevent some wave periods to reach the upper solar atmospheric layers (e.g., \citealt{1932hydr.book.....L, 2006ApJ...640.1153C, 2009ApJ...692.1211C, 2015MNRAS.450.3169P, 2020ApJ...896L...1M, 2024RSPTA.38230218K}). The acoustic cutoff frequency also plays an important role in helioseismology, which uses solar oscillations to determine the internal structure of the Sun (e.g., Basu 1987), and in asteroseismology, which deals with oscillations of different stars (e.g., \citealt{2019LRSP...16....4G}). Moreover, the cutoff has also been used to study free atmospheric oscillations of the Earth (\citealt{1998Sci...279.2089S}; Rhie \& Romanowicz 2004) and other planets (\cite{1998Natur.395..357K}), and acoustic oscillations of Jupiter (\citealt{1989ApJ...343..456D} Lee 1993).\\ 
The cutoff periods are fully determined the local physical conditions in the solar atmosphere, which means that the cutoffs are sensitive functions of the atmospheric height. They also depend on the physical conditions since different regions, such as sunspots, pores and faculae, exhibit differences in their cutoff value \citep{2009ApJ...692.1211C}.\\
The quiet sun (QS) has a network with a strong magnetic field, and an inter-network (IN) with either a weak field or practically magnetic field-free regions. The IN supports magnetoacoustic waves as well as purely magnetic modes such as Alfv\'en waves. Such waves are important means of transporting energy to higher layers from the convection zone and photosphere, and therefore, these waves are potential candidates for the required heating of the solar chromosphere and corona \citep{2015ApJ...806..170S}. Hence, the propagation of waves in IN is an key research topic to determine the role of different wave modes in the still not-fully-resolved problem of heating of the upper solar atmopsheric layers (e.g., \citealt{1982ApJ...253..367L, 1997ApJ...481..500C, 2004ApJ...617..623B, Kayshap2018}).\\
Finding the wave periods that can reach the upper layers of the solar atmosphere requires the determinantion of cutoff variations in the atmosphere.  This has been performed through observational, numerical, or theoretical studies. Various analytical formulas have been derived to determine local values of cutoff periods in the solar atmosphere
(e.g., \citealt{1966AnAp...29...55S, 1993A&A...273..671F, 2014AN....335.1043R}) that predict variations of the cutoffs with the atmospheric height. Typically, these studies require some unrealistic assumptions, such as imposing an isothermal atmosphere. In contrast, numerical simulations use more realistic models of the solar atmosphere that may account for temperature gradients, partial ionization, more complex magnetic field structures, and others. Variations of the cutoffs predicted by such numerical modeling can be directly compared to observations (e.g., \citealt{1978A&A....70..487U, 1997ApJ...481..500C,2012MNRAS.421..159F, 2019ApJ...882...32W, 2024RSPTA.38230218K}). \\ 
To date, there are only two reported observational attemtps 
to directly measure the stratification of the cutoff period from acoustic and magnetoacoustic waves in the IN  
(e.g., \citealt{2016ApJ...819L..23W, 2024ApJ...966..187S}). \cite{2024ApJ...966..187S} estimated the cutoff periods at three heights, namely, within the photosphere (i.e., Mn~{\sc i} 2801.907~{\AA}), chromosphere (i.e., Mg~{\sc ii} k 2796.35~{\AA}), and transition-region (i.e., Si~{\sc iv} 1393.77~{\AA}). Previously, \cite{2016ApJ...819L..23W} had probed the cutoff period at several heights within the photosphere/chromosphere, and reported variations in the acoustic cutoff period with height (see Figure 5 of \citealt{2016ApJ...819L..23W}). However, their comparison of the observed cutoff values to different theoretical formulas showed significant discrepancies between the theory and observations. A much better 
agreement between the observational cutoff and those retrieved from numerical simulations was reported by \cite{2019ApJ...882...32W} and \cite{2024RSPTA.38230218K}. \\
In this paper, the cutoff period at different heights in the solar atmosphere is evaluated by using Interface Region Imaging Spectrograph (IRIS) observations. Section~\ref{sect:cutoff_theor} describes the theory of cutoff periods and their comparisons with the observational estimations, and Section~\ref{sect:obs} is dedicated to the observations and data analysis. The results are described in Section~\ref{sec:results_obs} and they are discussed in the last Section.\\   
\section{Theories of cutoffs and their observational verifications} \label{sect:cutoff_theor}
\subsection{Acoustic and magnetoacoustic cutoff frequencies} As demonstrated first by Lamb (1909), the acoustic cutoff frequency arises naturally in stratified media, such as the Earth and solar atmospheres, and its value can be used to determine ranges of frequencies corresponding to propagating or evanescent waves (\citealt{Lamb1910, 1932hydr.book.....L}).  Originally, the acoustic cutoff frequency $\Omega_{ac} = c_s / 2 H_p $, where $c_s$ is the speed of sound and $H_p$ is a pressure scale height, was introduced by Lamb for an isothermal atmosphere with $c_s$ and $H_s$ being constant; this makes the cutoff to be a global quantity. \cite{Lamb1910} also considered a non-isothermal atmosphere with the temperature decreasing linearly with height, and studied the effects of such a uniform temperature gradient on $\Omega_{ac}$, which becomes then a local quantity in the atmosphere.\\
The temperature gradient considered by \cite{Lamb1910} was too simple to account for the temperature stratification observed in the solar atmosphere.  Some improvements were made by \cite{1984ARA&A..22..593D}, who modified Lamb's expression for $\Omega_{ac}$ by using $c_s$ and the density scale height ($H_{\rho}$) to be functions of the atmospheric height, and accounting for the derivative of $H_{\rho}$ with height. Further improvements were done by \cite{1993afd..conf..399G}, who derived $\Omega_{ac}$ by including the perturbation of gravitational potential and effects of spherical geometry. Two acoustic cutoff frequencies proposed by \cite{1998A&A...337..487S} directly involved first derivatives of $c_s$ with respect to the atmospheric height, and both are local quantities.  For the non-isothermal solar atmosphere with partially ionized plasma that is considered in this paper, the cutoff frequencies for ion and neutral acoustic waves were derived by \cite{2004ESASP.547....1R}.\\
Another method to determine the cutoff frequency for acoustic waves propagating in non-isothermal media without stratification was developed by \cite{2006PhRvE..73c6612M}. This method requires transformations of wave variables and the oscillation theorem to determine the turning point frequencies, and selecting the largest of them as the local cutoff. It was later generalized for non-isothermal atmospheres by \cite{2014AN....335.1043R}, who found that the acoustic cutoffs depend on the first and second derivatives of the sound speed with height. These studies did not account for the partially ionized plasma of the solar atmosphere.\\
The presence of magnetic fields also impacts the cutoff of the waves. The behavior of linear (magnetohydrodynamic) MHD waves propagating in a homogeneous medium with a uniform magnetic field of arbitrary directions is presently well-understood. There are three types of MHD waves: fast, slow, and Alfv\'en modes. In general, fast and slow MHD waves (also called magnetoacoustic waves) have both longitudinal and transverse components, but Alfv\'en waves are associated only with purely transverse motions (e.g., \citealt{1982ApJ...253..367L}; Thomas 1983, Campos 1986). In regions where the Alfv\'en speed exceeds the sound speed, slow magnetoacoustic waves behave like sound waves but are confined to propagate along magnetic field lines. Their cutoff is reduced by a factor of $\cos\theta$ (where $\theta$ is the magnetic field inclination from the vertical) since gravity is effectively reduced along inclined field lines \citep{1977A&A....55..239B, 2006ApJ...648L.151J}. In addition, numerous works have also reported cutoff frequencies for magnetoacoustic and Alf\'en waves (e.g., \citealt{1982ApJ...256..761R, 1990ApJ...351..287M, 1991A&A...250..235F, 1993A&A...273..671F, 1993ApJ...409..450S, 1995ApJ...452..434M, 1997LNP...489...75R, 2003A&A...409.1085N, 2006PhRvE..73c6612M, 2007ApJ...659..650M, 2010A&A...518A..37M, 2018A&A...617A..39F, 2020Ap&SS.365..139R, 2020ApJ...896L...1M,2022Ap&SS.367..111M}). \\
Among these many papers, \cite{1993ApJ...409..450S} provides the most relevant estimation of the cutoff frequencies for the waves analyzed in our observational work. This study developed the analytical expression for the local cutoff frequency of fast magnetoacoustic waves in an isothermal atmosphere embedded in a uniform horizontal magnetic field. These assumptions are inconsistent with the solar data (e.g., \citealt{2004soas.book.....F}), which show highly inhomogeneous structures in the observable part of the solar atmosphere \citep[e.g.,][]{1993SSRv...63....1S}. Therefore, a quantitative comparison between studies of MHD waves propagation in such uniform media and actual observational measurements is out of the scope of this paper. Also, our empirical estimation of the cutoff is based on the detection of the frequency at which a phase shift different from zero between two velocity signals is detected. Since radiative losses shift the phase difference spectra, our methodology will provide a ``pseudo-cutoff'' rather than the theoretical cutoff value \citep[e.g.,][]{2006ApJ...640.1153C}. In the following, we will refer to this effective cutoff.
\subsection{Observations of the cutoff stratification in the solar atmosphere}
The first direct detection of the acoustic cutoff frequency and its variations with height in the quiet regions of the solar upper photosphere and lower chromosphere was reported by \cite{2016ApJ...819L..23W}, who measured nine selected spectral lines by using the Vacuum Tower Telescope (VTT) located at Tenerife.  \cite{2016ApJ...819L..23W} compared the obtained data to the theoretical predictions made by the formulas for $\Omega_{ac}$ derived by \cite{1984ARA&A..22..593D}, \cite{1998A&A...337..487S}, and \cite{2006PhRvE..73c6612M} and found significant discrepancies between their data and the theoretical predictions. \\
In the follow-up work, \cite{2024ApJ...966..187S} used the data collected by the Interface Region Imaging Spectrograph (IRIS) to establish the existence of acoustic waves and variations of the acoustic cutoff frequency in the solar atmosphere.  Three spectral lines (one in the solar photosphere, one in the solar chromosphere and one in the transition region) were observed, and clear variations of the acoustic cutoff in these atmospheric layers were detected. The obtained data spans a much larger fraction of the solar atmosphere than those reported by \cite{2016ApJ...819L..23W}, and again, no known theoretical formula for the acoustic cutoffs could explain the data.  In the case of strongly magnetized atmospheres, \cite{2018A&A...617A..39F} also reported changes with height of the cutoff in sunspots from infrared observations acquired with the GREGOR telescope.\\
\section{Observation and Data Analysis}\label{sect:obs}
The QS, which is located between two ARs, was observed by IRIS on November 16$^{th}$, 2013. The QS observations are performed in the sit-n-stare mode to capture the temporal evolution along the slit of the intensity spectra in the far-ultraviolet (1331.7~{\AA}{--}1358.4~{\AA} and  1389.0~{\AA}{--}1407.0~{\AA}) and near-ultraviolet (2782.7~{\AA}{--}2835.1~{\AA}). The temporal cadence of these observations is 16.7 s. In addition to spectroscopic observations, IRIS also captures various slit-jaw images (SJIs) of QS.
\begin{figure}[ht!]
\includegraphics[trim = 2.0cm 0.0cm .0cm 0.0cm, scale=0.90]{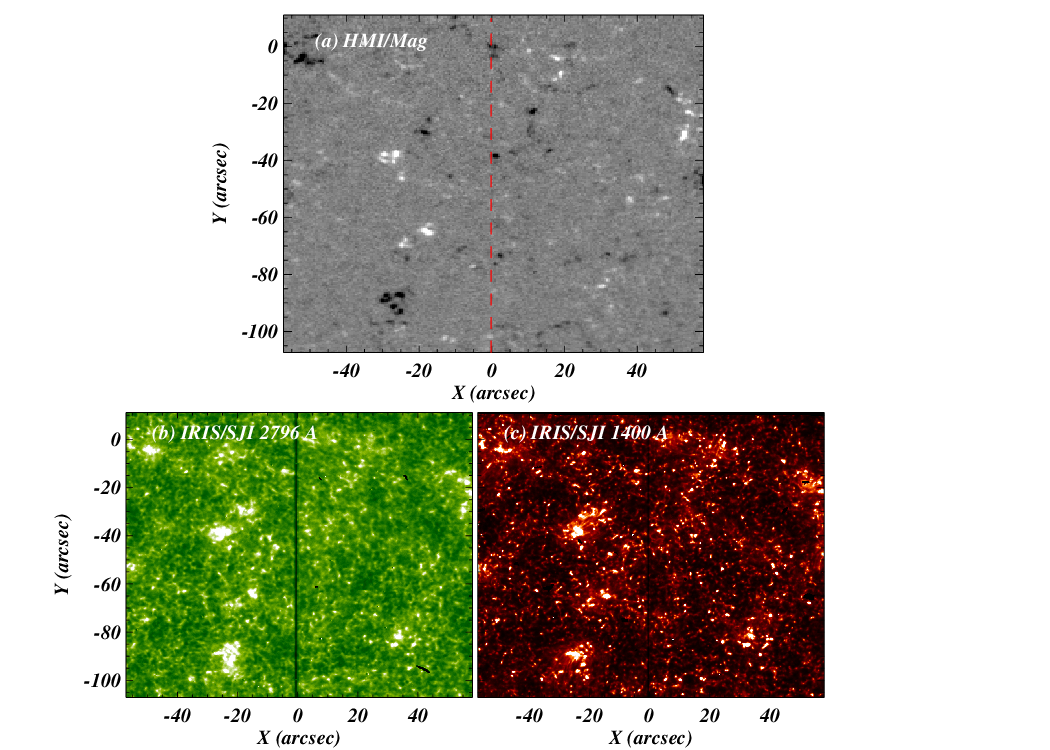}
\caption{LOS magnetogram (panel a),  chromospheric (i.e., IRIS/SJI~2796~{\AA}, panel b), and transition-region intensity (i.e., IRIS/SJI~1400~{\AA}, panel c) of the QS region. The black vertical lines in panels (b) and (c) show the IRIS/slit position.}  
\label{fig:ref_fig}
\end{figure}
Figure~\ref{fig:ref_fig}(a) shows the line-of-sight (LOS) magnetogram, observed by the Helioseismic and Magnetic Imager \citep[HMI,][]{2012SoPh..275..207S} onboard Solar Dynamics Observatory (SDO), of the QS. The LOS magnetogram shows the existence of weak magnetic fields in the region of interest (ROI). The very weak magnetic field at the location of the IRIS slit (red-dashed line in Fig.~\ref{fig:ref_fig}(a)) indicates that the ROI corresponds to a region of IN. This is also confirmed by the low intensity seen at the slit position in the chromosphere (IRIS/SJI 2796~{\AA}, Fig.~\ref{fig:ref_fig}(b)) and transition region (IRIS/SJI 1400~{\AA}, Fig.~\ref{fig:ref_fig}(c)). This observation has previously been employed by \cite{2025ApJ...987..161K} to estimate the dominant period at multiple heights.\\  
Table~\ref{tab:table1} presents the details (i.e., ion, wavelength,  formation height$\pm$uncertainty in the formation height, and the layer of the solar atmosphere) of six photospheric absorption lines and one chromospheric emission spectral line that are utilized to estimate the cutoff period at multiple heights in the photosphere and chromosphere. The heights of these spectral lines range from 0.17 Mm to 2.2 Mm within the solar atmosphere. 
\begin{table}[]
\caption{The used spectral lines along with their wavelength, formation height, and uncertainties in the formation height.}
    \centering
    \hspace{0.0cm}
    \begin{tabular}{|c|c|c|c|c|}
       \hline
    Sr. No. & Spectral Line & Wavelength (A) & Formation Height (Mm) &  Formation Layer \\
    \hline
    1 & Ni~{\sc i} & 2815.179 &  0.17$\pm$0.03 &  photosphere \\ 
    \hline
    2 & Fe~{\sc i} & 2792.327 &  0.38$\pm$ 0.03 & photosphere \\
    \hline
    3 & Fe~{\sc i} & 2793.223 & 0.50$\pm$0.04 &  photosphere \\
    \hline
     4 & Ni~{\sc i}  & 2799.347 &  0.68$\pm$0.07 & photosphere \\
    \hline
    5 & Fe~{\sc i}  & 2814.114 &  0.76$\pm$0.11 &  photosphere \\
    \hline
    6 & Mn~{\sc i}  & 2801.907 &  0.83$\pm$0.10 & photosphere\\
    \hline
    7 & Mg~{\sc ii} k2r & 2796.35 &  1.40$\pm$0.30 &  chromosphere\\
    \hline
    8 & Mg~{\sc ii} k3 & 2796.35 &  1.90$\pm$0.10 & chromosphere\\
    \hline
    \end{tabular}
    \label{tab:table1}
\end{table}
The information from Table~\ref{tab:table1} about the photospheric absorption lines is obtained from the IRIS website\footnote[1]{https://iris.lmsal.com/itn39/photospheric.html \label{fn_1}}. The Mg~{\sc ii} k 2796.35~{\AA} line is an optically thick line, and forms over a large height range \citep{1981ApJS...45..635V}. The formation heights of k2r, k2v, and k3 are 1.40, 1.40, and 1.90 Mm, respectively (\citealt{1981ApJS...45..635V}). However, \cite{2013ApJ...772...90L} showed that Mg~{\sc ii} k2v forms slightly higher than Mg~{\sc ii} k2r.\\
The methodology for the fitting of photospheric absorption lines is described in the ITN 39 (see footnote~\ref{fn_1}). The profile of each absorption line from each spatial location and time is fitted using a Gaussian fit, and the intensity, centroids, and sigma are estimated. Further, using the rest wavelength, the centroid is converted into Doppler velocity. Unlike the photospheric absorption lines, we cannot apply a single Gaussian fit on the Mg~{\sc ii} because it is an optically thick line. Instead, we have used the routine iris$\_$get$\_$mg$\_$features.pro from \cite{2013ApJ...772...90L} to characterize the Mg~{\sc ii} line and estimate the Doppler velocities from Mg~{\sc ii} k2r and k3. Finally, for all spectral lines in Table(~\ref{tab:table1}), Doppler velocity-time series (hereafter DTS) have been constructed for all the 700 spatial locations along the slit. As an example, Figure~\ref{fig:dts} shows the DTS of Ni~{\sc i} 2815.18~{\AA} and Fe~{\sc i} 2792.33~{\AA} at a randomly selected spatial location.
\begin{figure}[ht!]
\includegraphics[trim = 2.7cm 1.5cm 3.0cm 2.0cm,scale=1.0]{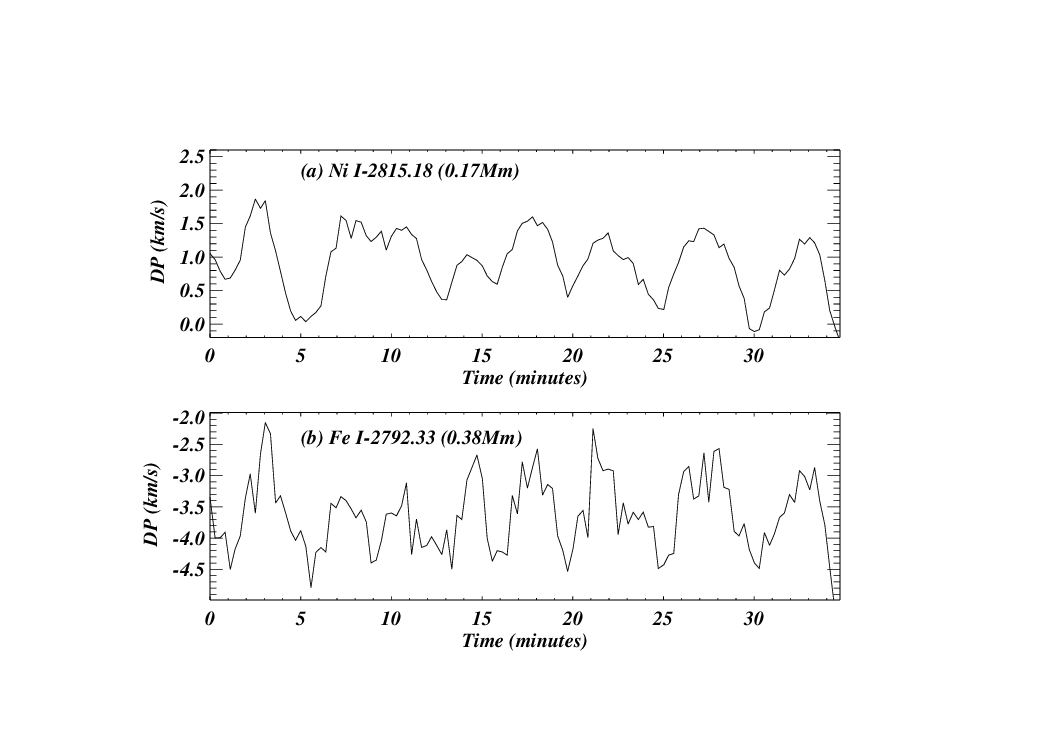}
\caption{Temporal evolution of the Doppler velocity of Ni~{\sc i} 2815.18~{\AA} (panel a) and Fe~{\sc i} 2792.33~{\AA} (panel b) at one particular spatial location.} 
\label{fig:dts}
\end{figure}
\subsection{Phase Differences and wavelet analysis}
\begin{figure}[ht!]
\includegraphics[trim = 1.0cm 0.0cm 2.0cm 0.0cm,scale=0.82]{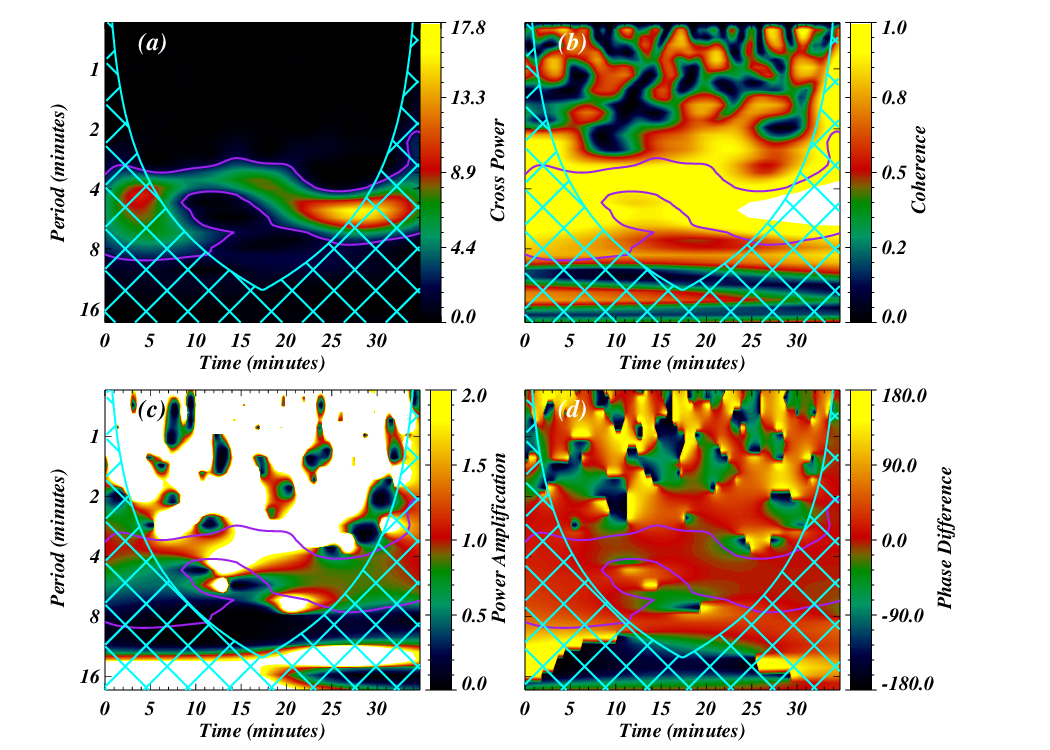}
\caption{Wavelet analysis of the two DTS shown in Figure~\ref{fig:dts}, including the cross wavelet power (panel a), coherence (panel b), amplification (panel c) and phase difference (panel d) maps.  The purple contour encloses the regions with cross power above 95\% significance level in panel (a), and the same contour is drawn in other panels. The  cyan cross-hatched area outlines the cone-of-influence (COI).} 
\label{fig:cross_wavelet}
\end{figure}

Cross-wavelet analysis has been used to compute the cross power, coherence, and phase difference between two time series originating from two different heights (e.g., \citealt{2004ApJ...617..623B, 2017ApJS..229...10J, Kayshap2018, 2022MNRAS.517..458S}). Here, the cross-wavelet analysis is performed between two DTS originating from the two closest heights, as indicated in Table~\ref{tab:table2}.\\
\begin{table}[]
\caption{Pairs of spectral lines employed for the cross-wavelet analysis}
    \centering
    \hspace{0.0cm}
    \begin{tabular}{|c|c|c|}
     \hline
     Pair. No. & Spectral Lines & Probed Heights (Mm)  \\
     \hline
     1 & Ni~{\sc i}~2815.179~{\AA}{--} Fe~{\sc i}~2792.327~{\AA} & 0.17{--}0.38 \\ 
     \hline
     2 & Fe~{\sc i}~2792.327~{\AA}{--}Fe~{\sc i}~2793.223~{\AA} & 0.38{--}0.50 \\
     \hline
     3 & Fe~{\sc i}~2793.223~{\AA}{--}Ni~{\sc i}~2799.347~{\AA} & 0.50{--}0.68 \\
     \hline
     4 & Ni~{\sc i}~2799.347~{\AA}{--}Fe~{\sc i}~2814.114~{\AA} & 0.68{--}0.76 \\
     \hline
     5 & Fe~{\sc i}~2814.114~{\AA}{--}Mn~{\sc i} 2801.907~{\AA}  & 0.76{--}0.83\\
     \hline
     6 & Mn~{\sc i} 2801.907~{\AA}{--}Mg~{\sc ii} 2796.35~{\AA}~k2r &  0.83{--} 1.4\\
     \hline
     7 & Mg~{\sc ii} 2796.35~{\AA}~k2r{--}Mg~{\sc ii} 2796.35~{\AA}~k3 &  1.4{--}1.9\\
    \hline
    \end{tabular}
    \label{tab:table2}
\end{table}
Figure~\ref{fig:cross_wavelet} illustrates the wavelet analysis of the two DTS shown in in Figure~\ref{fig:dts}, i.e., DTS originating from the heights of 0.17 Mm (i.e., Ni~{\sc i} 2815.179~{\AA}) and 0.38 Mm (i.e., Fe~{\sc i}~2792.327~{\AA}). Significant cross-power lies in a period range of 3.0 to 7.0 minutes (Figure~\ref{fig:cross_wavelet}(a)). The regions with cross power above a 95\% confidence level are delimited  by purple contour lines in panel (a), and the same contour is drawn in other panels. These regions with high confidence exhibit a higher coherence between both signals (Figure~\ref{fig:cross_wavelet}(b)). Also, they show a positive phase difference, which indicates that the upper velocity signal (measured with the Fe~{\sc i}~2792.327~{\AA} line) is lagging the velocity from the lower layer (Ni~{\sc i} 2815.179~{\AA}). Thus, the phase difference is consistent with upward wave propagation. The power amplification (panel c), defined as the ratio between the power at the upper height (0.38 Mm) and at a lower height (0.17 Mm), mostly exhibits a power enhancement with height.\\
In this work, we have evaluated the phase difference and amplification spectra between the set of line pairs indicated in Table \ref{tab:table2}. Our goal is to assess the cutoff frequency from the examination of these spectra (e.g., \citealt{2006ApJ...640.1153C, 2009ApJ...692.1211C, 2016ApJ...819L..23W, Kayshap2018, 2024ApJ...966..187S}). If the phase difference ($\delta$$\phi$) is equal/close to zero, those particular waves are standing/evanescent waves. In contrast, $\delta$$\phi$ $\neq$ 0 indicates wave propagation. The phase difference spectrum between the velocities at two different heights generally exhibit a progressive change from $\delta$$\phi$ around zero at the low-frequency evanescent waves to positive values at higher frequencies, indicating upward wave propagation. The frequency at which the phase shift depart from zero is defined as the cutoff value (e.g., \citealt{2009ApJ...692.1211C, 2018A&A...617A..39F}).\\
We have computed the mean values of the phase difference, amplification, and coherence spectra by averaging the measurements from all the spatial locations along the slit and times. In this computation, only the regions with reliable values have been included. These regions need to satisfy the following conditions: (1) be outside of the COI, (2) lie within the 95\% confidence level, and (3) have a coherence value equal to or greater than 0.7.  The coherence varies from 0 (no coherence at all in two time series) to 1 (perfect coherence between two time series). However, usually, the coherence value exceeding 0.7 reflects the strong correlation between two time-series (e.g., \citealt{2017ApJS..229...10J}). Therefore, we have used this threshold value of coherence (i.e., 0.7) in this analysis, as this criterion helps to pick the region which are strongly correlated between two DTS. The values of phase difference, amplification, and coherence that fulfill these criteria are averaged over time for each spatial location. Now, we have 700 spatial locations of each parameter (i.e., phase difference, amplification, and coherence) as a function of wave period. Finally, all the spatial locations were averaged. The resultant spectra are illustrated in Figures ~\ref{fig:cf1} to ~\ref{fig:cf7} for all the line pairs, where we have replaced the wave period by the wave frequency.    
\section{Results} \label{sec:results_obs}
\begin{figure}[ht!]
\includegraphics[trim = 1.0cm 1.5cm 1.0cm 1.0cm,scale=0.74]{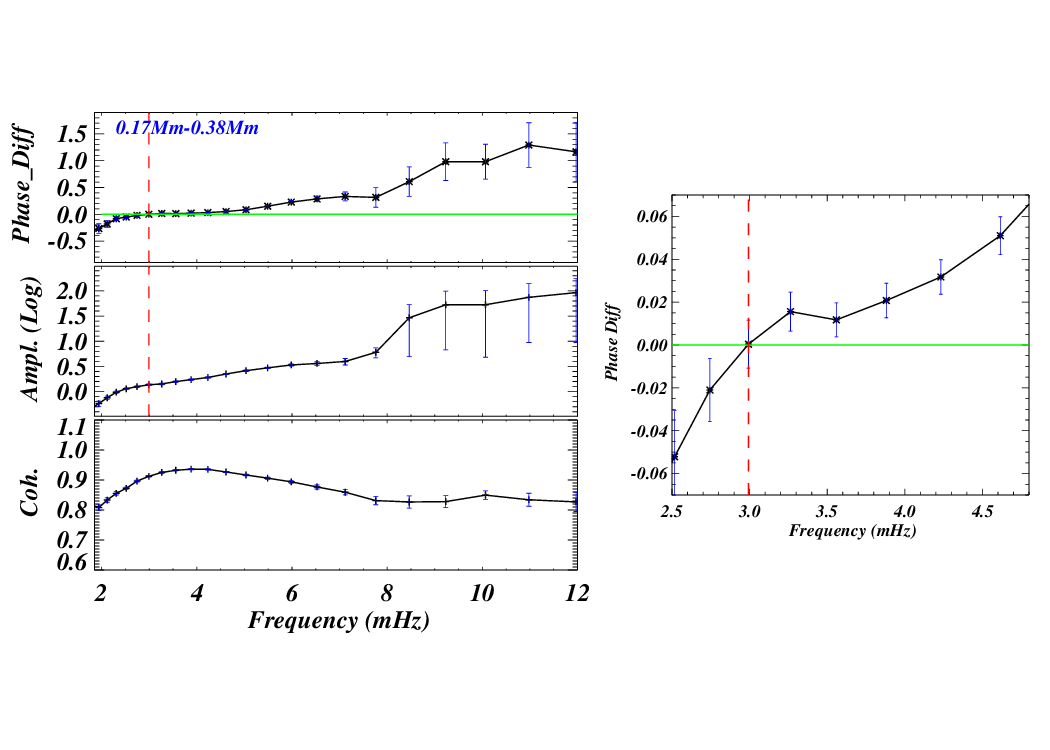}
\caption{Figure shows the mean $\delta$$\phi$ (top-left panel), power amplification (middle-left panel), and coherence (bottom-left panel) with frequency for the first line pair, i.e., 0.17{--}0.38 Mm. The standard errors are shown in blue in each panel. The horizontal green line lies at zero value of $\delta$$\phi$. The $\delta$$\phi$ for a small frequency range (i.e., from 2.5 to 4.8 mHz) is shown in the right panel to estimate the cutoff frequency, i.e., frequency close to zero $\delta$$\phi$. The vertical red-dashed line is placed at the cutoff frequency. The same red dashed line is shown in panels (a) and (b).}  
\label{fig:cf1}
\end{figure}
\begin{figure}[ht!]
\includegraphics[trim = 1.0cm 1.5cm 1.0cm 1.0cm,scale=0.74]{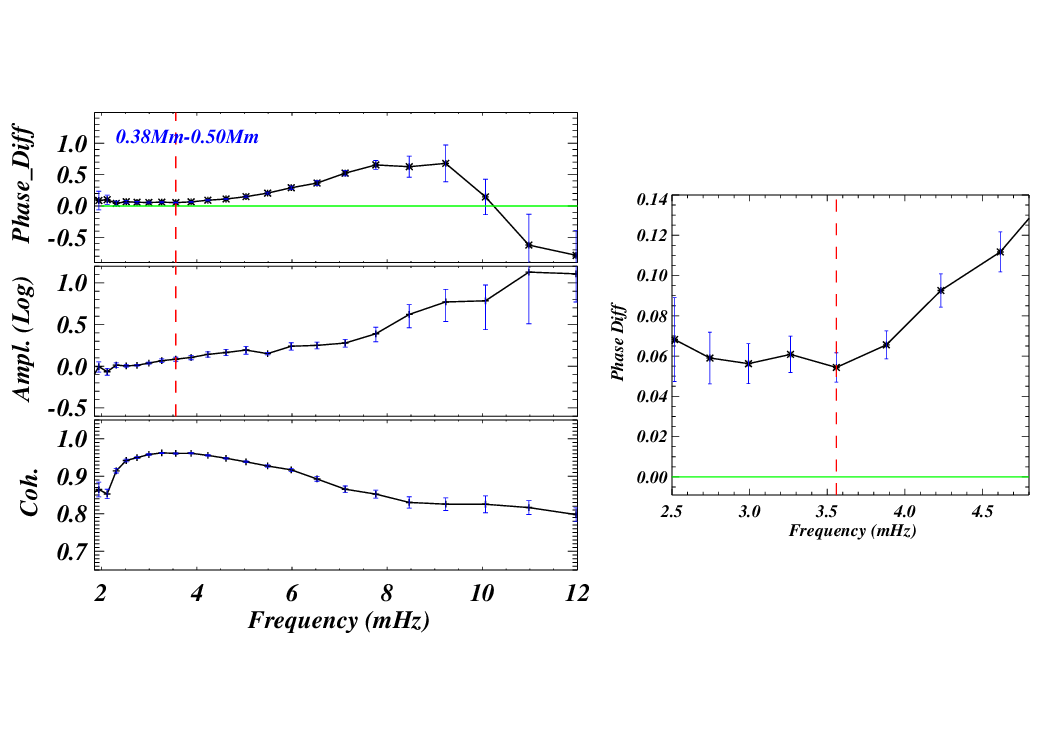}
\caption{Same as Figure~\ref{fig:cf1} but for second line pair, i.e., 0.38{--}0.50 Mm.} 
\label{fig:cf2}
\end{figure}
Figure~\ref{fig:cf1} shows the averaged (over time and spatial locations) $\delta$$\phi$ (top-left panel), power amplification (middle-left panel), and coherence (bottom-left panel) for the lowest height pair (i.e., 0.17 to 0.38 Mm). The error bars shown in blue are the standard errors. As a reference, a green horizontal line at zero $\delta$$\phi$ is included in the plot. Initially, the $\delta$$\phi$ is negative at a frequency of 2 mHz, and further, the $\delta$$\phi$ moves towards zero $\delta$$\phi$. At higher frequencies $\delta$$\phi$ becomes positive. A zoomed view of $\delta$$\phi$ in a small frequency range from 2.5 to 4.8 mHz is shown in the right panel to better visualize the frequency at which $\delta$$\phi$ first becomes positive. A value of 3.0 mHz, indicated by a vertical dashed line, as been selected as the cutoff frequency in the low photosphere probed by these spectral lines.\\
\begin{figure}[ht!]
\includegraphics[trim = 1.0cm 1.5cm 1.0cm 1.0cm,scale=0.74]{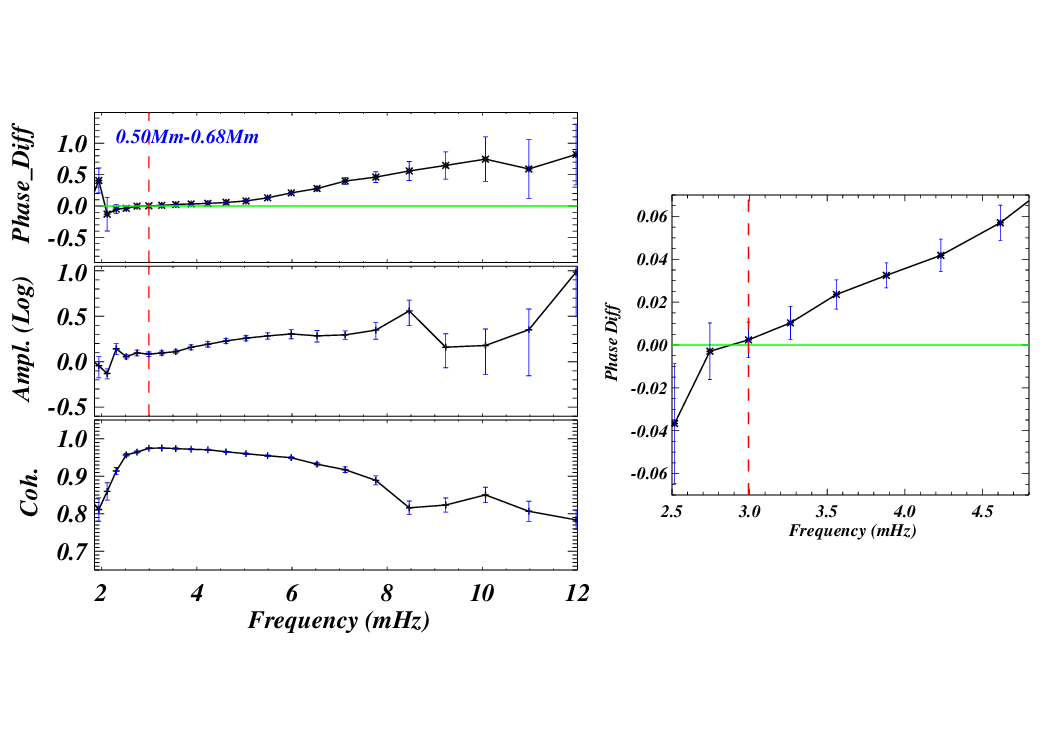}
\caption{Same as Figure~\ref{fig:cf1} but for third line pair, i.e., 0.50{--}0.68 Mm}  
\label{fig:cf3}
\end{figure}
\begin{figure}[ht!]
\includegraphics[trim = 1.0cm 1.5cm 1.0cm 1.0cm,scale=0.75]{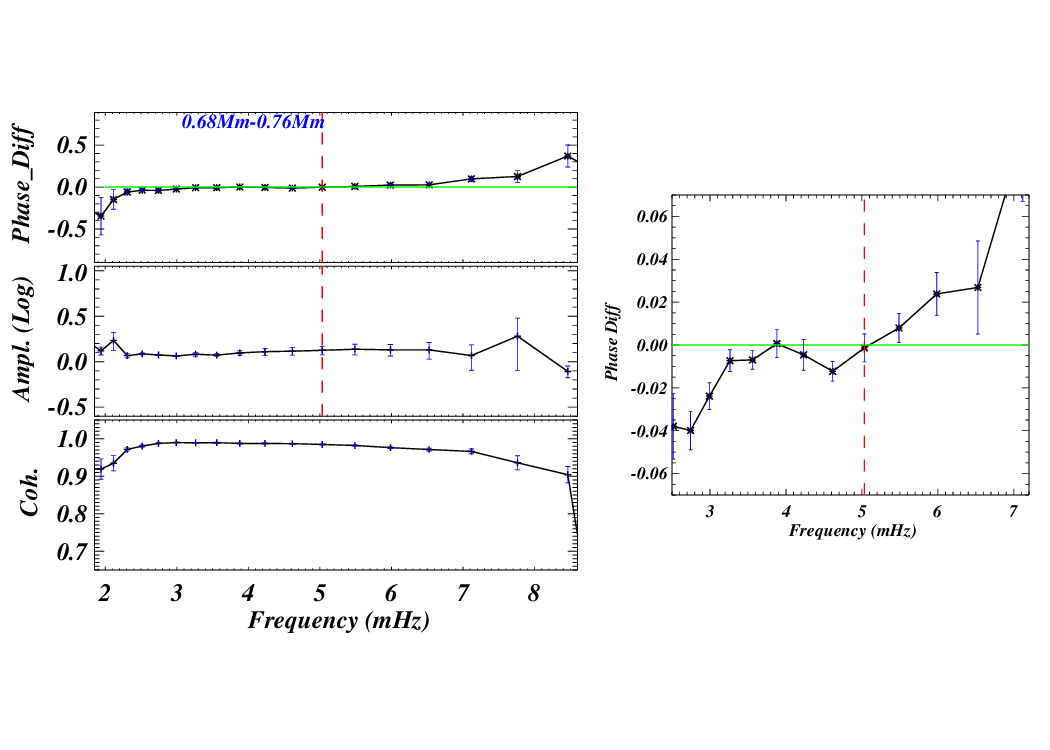}

\caption{Same as Figure~\ref{fig:cf1} but for fourth line pair, i.e., 0.68{--}0.76 Mm.}  
\label{fig:cf4}
\end{figure}

\begin{figure}[ht!]
\includegraphics[trim = 1.0cm 1.5cm 1.0cm 1.0cm,scale=0.75]{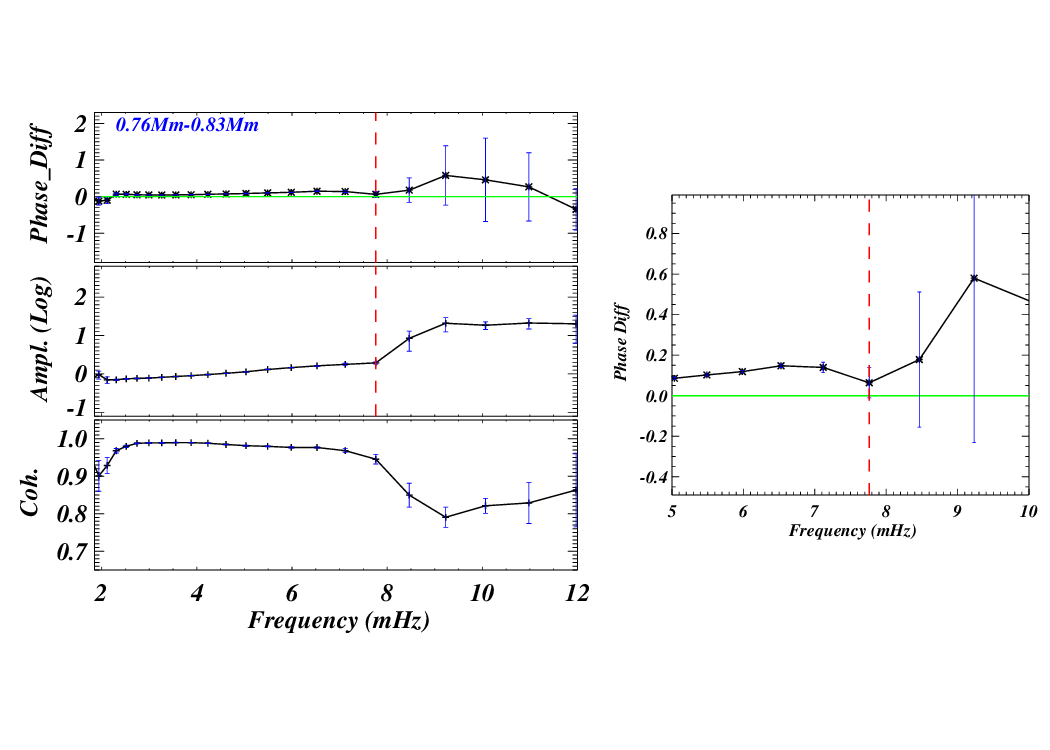}

\caption{Same as Figure~\ref{fig:cf1} but for fourth line pair, i.e., 0.76{--}0.83 Mm}  
\label{fig:cf5}
\end{figure}

\begin{figure}[ht!]
\includegraphics[trim = 1.0cm 1.5cm 1.0cm 1.0cm,scale=0.75]{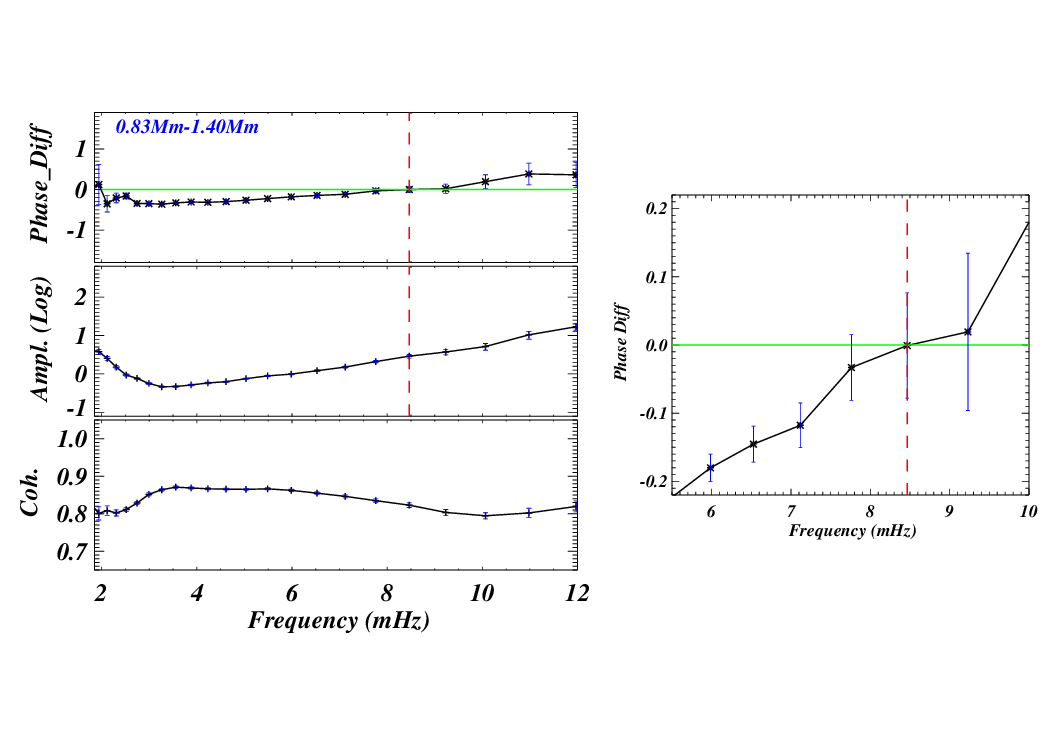}

\caption{Same as Figure~\ref{fig:cf1} but for fourth line pair, i.e.,  0.83{--}1.40 Mm}
\label{fig:cf6}
\end{figure}

\begin{figure}[ht!]
\includegraphics[trim = 1.0cm 1.5cm 1.0cm 1.0cm,scale=0.75]{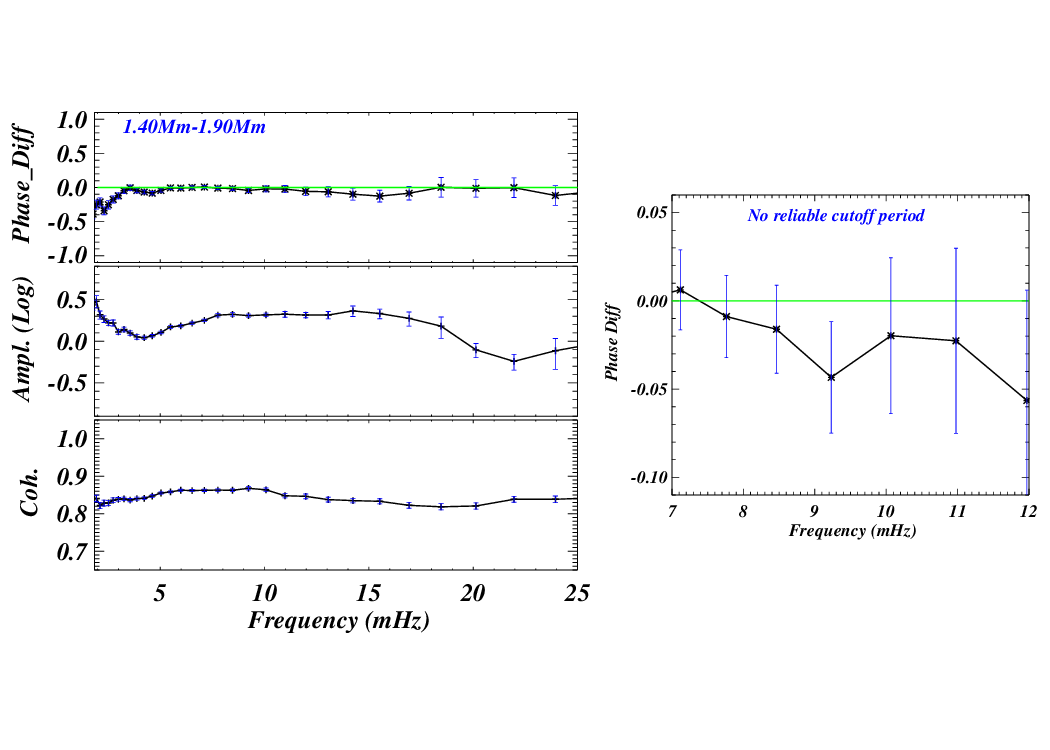}

\caption{Same as Figure~\ref{fig:cf1} but for fourth line pair, i.e.,  1.4{--}1.9 Mm}  
\label{fig:cf7}
\end{figure}
The same analysis and estimation of the cutoff frequency have been applied to the rest of the line pairs and is shown in Figures~\ref{fig:cf2} (0.38 Mm{--}0.50 Mm), ~\ref{fig:cf3} (0.50 Mm{--}0.68 Mm), ~\ref{fig:cf4} (0.68 Mm{--}0.76 Mm), ~\ref{fig:cf5} (0.76 Mm{--}0.83 Mm), ~\ref{fig:cf6} (0.83 Mm{--}1.20 Mm), and ~\ref{fig:cf7} (1.20 Mm{--}2.20 Mm). The vertical red-dashed line in all these Figures is located at the respective cutoff frequency. Our results show that most of the cases are consistent with the general picture of evanescent low frequency waves ($\delta$$\phi\sim0$) which progressively turn into upward propagating waves as higher frequencies are considered. Wave propagation is also generally accompanied by an increase in the amplification between the sampled heights, as expected from waves propagation in a gravitationally stratified atmosphere. \\
A few cases depart from this picture. The phase difference between velocity signals at 0.38 Mm and 0.50 Mm (Figure~\ref{fig:cf2}) exhibits an continuous increase up to 9 mHz. At higher frequencies, $\delta$$\phi\sim0$ suddenly decreases. The same phenomenon is found for the pairs (0.68 Mm{--}0.76 Mm, Figure~\ref{fig:cf4}), and (0.76 Mm{--}0.83 Mm,  Figure~\ref{fig:cf5}), where the phase shift also starts to decrease for frequencies above 8.5-9 mHz. Similar roll-offs have been reported by Fleck et al. (2021), Zhao et al. (2022), and Chaturmutha et al. (2024). In these works, they are generally found for low height separations. Our results are consistent with those previous measurements since we only obtain them when the height difference is below 120 km. The cases with larger separations (210 km in Figure~\ref{fig:cf1}, 180 km in Figure ~\ref{fig:cf3}, 370 km in Figure~\ref{fig:cf6}) do not present roll-offs or only show slight indications of them. They are probably caused by the temporal undersampling of the data (Chaturmutha et al., 2024).\\
Figure~\ref{fig:cf7} illustrates the results from the oscillations between the two highest lines (i.e., 1.2 Mm to 2.2 Mm). It evaluates the wave propagation between the low and high chromosphere. In the range between 3 and 25 mHz, all the $\delta$$\phi$ values are very close to zero, showing that waves at these two distant layers are oscillating in phase. They are consistent with standing oscillations, produced by waves trapped in a chromospheric resonant cavity. Since our cutoff estimation is based on the detection of phase differences, we cannot retrieve a cutoff value for these heights.
\begin{figure}[ht!]
\includegraphics[trim = 2.0cm 0.0cm 3.0cm 0.0cm,scale=0.90]{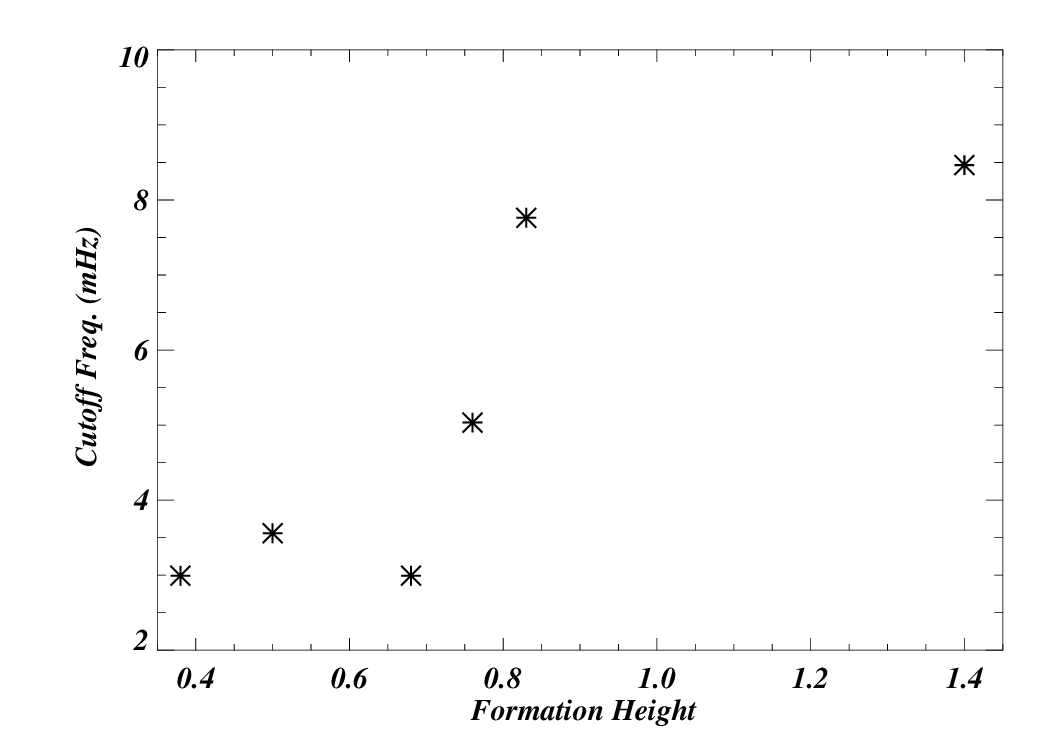}

\caption{Cutoff frequency as a function of the atmospheric height. 
\label{fig:cutoff_freq}}
\end{figure}
The cutoff frequencies derived from the examination of Figures ~\ref{fig:cf1} to ~\ref{fig:cf6} are plotted as a function of height in Figure~\ref{fig:cutoff_freq}. For each case, the inferred cutoff value has been assigned to the lower height of the pair of lines probed. However, we note that in our analysis we are not determining the cutoff at that specific location. Instead, we are evaluating the lower frequency waves that can propagate at least in some region between the two probed layers. The cutoff stratification exhibits an increase from 3 mHz at 0.38 Mm to 8.5 mHz at 1.2 Mm. This increase is not progressive, but a sudden jump in the cutoff value is found at around 0.8 Mm, where a striking increment from 3 to 8 mHz takes place over a 0.15 Mm span.
\section{Discussion and Conclusions} \label{sec:discussion}   
Wave propagation and dissipation of wave energy in different parts of the solar atmosphere have been an important field of study for many years, with the primary goal being the local atmospheric heating (e.g. \citealt{2015LRSP...12....6K, 2015RSPTA.37340261A, 2023LRSP...20....1J}). Now, it has become a consensus that 3-minute wave periods in the QS propagate into the higher layers from the photosphere. It has also been demonstrated that the cutoff period in the chromosphere is around 4 minutes (e.g., \citealt{1979ApJ...231..570L, 1982ApJ...253..367L, 2006ApJ...640.1153C, 2009ApJ...692.1211C, Kayshap2018, Kayshap2020, 2024ApJ...966..187S}).\\

This means that the 5-minute periods that dominate in the solar photosphere generally do not reach the chromosphere since they become evanescent. However, these evanescent 5-minute periods can reach into chromosphere/TR though several physical processes. On of them is the propagation along inclined magnetic fields, as suggested by several works \citep{1977A&A....55..239B, 2006ApJ...648L.151J, 2005ApJ...624L..61D}. This way, in different magnetized regions of the solar atmosphere (e.g., networks and solar plages), 5-minute period waves are allowed into the chromosphere (e.g., \citealt{Kayshap2020, 2026AdSpR...xxx...xxG}). \cite{2009ApJ...692.1211C} estimated the cutoff frequencies in sunspots, pores, and facular regions at the chromospheric heights. They reported that the atmospheric cutoff frequency is around 4 mHz in sunspot-like structures, while the cutoff frequency drops below 2 mHz for facular regions, i.e., longer-period waves are allowed to reach in the chromosphere from the photosphere. Radiative losses also contribute to the propagation of 5-minute period waves into the solar chromosphere. When the radiative relaxation time is small enough, the effective cutoff frequency is reduced, allowing the propagation of long period waves (\citealt{2006ApJ...640.1153C, 2008ApJ...676L..85K, 2020A&A...640A...4F}). \\

The above brief description of the previously obtained results focuses only on the conditions that allow the photospheric long-period waves to reach the chromosphere. However, in their travel to upper atmospheric layers, waves also encounter variations in the cutoff frequency since it is a quantity that depends of the local conditions. The main goal of this paper is to observationally determine this theoretically predicted stratification (see Section~\ref{sect:cutoff_theor}) and provide new independent empirical results. \\

\cite{2016ApJ...819L..23W} inferred the variations of the cutoff frequency in the QS at different heights from the photosphere to the chromosphere. Their observational results were compared against cutoff frequencies predicted by various theoretical works (e.g., \citealt{Lamb1910, 1984ARA&A..22..593D, 1998A&A...337..487S, 2006PhRvE..73c6612M}), but they found a large difference between the observational and theoretical cutoff frequencies. In contrast, the numerical simulations from \citealt{2016ApJ...827...37M} showed that the numerically estimated dominant frequency (i.e., related to the cutoff frequency) was pretty close to the observational cutoff frequency estimated by \cite{2016ApJ...819L..23W}. Also, these observational estimations have been compared with various numerical simulations based on single and double fluids (e.g., \citealt{2019MNRAS.482.3244K, 2019ApJ...882...32W, 2022A&A...668A..32N, 2024RSPTA.38230218K}) which have established good agreements between the observations and simulations.\\

Our study shares the same goals and target of \cite{2016ApJ...819L..23W}, but a direct comparison is hindered due to some methodogical differences. We have calculated the phase differences between pairs of spectral lines at close heights, rather than limiting the analysis to the line with the highest formation height compared to the others. This approach enables us to examine more localized atmospheric layers and to constrain the inferred cutoff frequency to a narrower and better-defined height range. Our methodology is similar to \cite{2018A&A...617A..39F}, who estimated the cutoff frequency at multiple heights in the umbral atmosphere. \\

The present work reports a comprehensive estimation of the cutoff frequencies using IRIS observations, i.e., we estimated the cutoff frequency at six heights between the photosphere and the chromosphere. We find an increase in the cutoff frequency with height from around 3 mHz (at 0.38 Mm) to around 8.5 mHz (at 1.20 Mm). We also find indications of standing waves at higher atmospheric layers between the low and high chromosphere. IRIS observations have also been employed to obtain independent estimations of the cutoff frequency. \cite{Kayshap2018} inferred a cutoff frequency around 4 mHz at the chromosphere. More recently, \cite{2024ApJ...966..187S} estimated the cutoff frequency in the QS at the photosphere (Fe~{\sc i} 2799.972{--}Mn~{\sc i} 2801.907), (2) chromosphere (Fe~{\sc i} 2799.972{--}Mg~{\sc ii} h 2803.52), and (3) transition-region (Fe~{\sc i} 2799.972{--}Si~{\sc iv} 1393.755). They found that all the waves from 3 to 8 mHz propagate from the formation height of the Fe~{\sc i} line to the Mn~{\sc i} line within the photosphere. They also found a reduction in the cutoff from 4.7 mHz in the chromosphere to 3.14 mHz in the transition region. \\

Our results agree with previous works by showing clear evidence of the stratification of the cutoff in the solar atmosphere, but also exhibit some quantitative differences. The main difference is the relatively high cutoff value that we have inferred in the low chromosphere. Several reasons can lead to this mismatch. First, we are estimating the cutoff from the phase differences between the velocities from pairs of lines that form at very close atmospheric heights. Although this provides a much better constraint in the probed height, it also supposes an observing challenge since it requires the measurement of very small phase differences. This cutoff value should be regarded as an upper bound of the actual cutoff. Waves at lower frequencies may still propagate, but their small phase difference, caused by the minimal separation in formation height between the two lines, could make them indistinguishable within the uncertainties of our measurements. Second, higher chromospheric layers host standing oscillations. Possibly, at the low chromosphere, where we find high cutoff values, propagating waves co-exist with standing waves whose phase difference is constant with height. The superposition of the wave signatures from propagating and standing waves will tend to reduce the average phase shift, leading to an increment in the estimated cutoff frequency. More analyses are required to find a convergence in the cutoffs inferred from observations, since they are a fundamental input for the verification of various theoretical models of wave propagation in solar/stellar atmospheres.  

\begin{acks}
 We acknowledge the constructive comments/feedback on the manuscript. IRIS is a NASA small explorer mission developed and operated by LMSAL, with mission operations executed at NASA Ames Research Center, and major contributions to downlink communications funded by ESA and the Norwegian Space Centre. TF acknowledges grants PID2021-127487NB-I00, PID2024-156538NB-I00, CNS2023-145233, and RYC2020-030307-I funded by MCIN/AEI/10.13039/501100011033. KM's research was conducted within the framework of the project from the Polish National Foundation (NCN) grant No. 2025/57/B/ST9/00114.
\end{acks}





\section{Additional statments}

\begin{authorcontribution}

\end{authorcontribution}

\begin{fundinginformation}
Information about fundings ...
\end{fundinginformation}

\begin{dataavailability}
Information about available data ...
\end{dataavailability}

\begin{materialsavailability}
Information about available material ...
\end{materialsavailability}

\begin{codeavailability}
Information about available code ...
\end{codeavailability}

\begin{ethics}
\begin{conflict}
The authors declare that they have no conflicts of interest ....
\end{conflict}
\end{ethics}


\IfFileExists{\jobname.bbl}{} {\typeout{}
\typeout{****************************************************}
\typeout{****************************************************}
\typeout{** Please run "bibtex \jobname" to obtain} \typeout{**
the bibliography and then re-run LaTeX} \typeout{** twice to fix
the references !}
\typeout{****************************************************}
\typeout{****************************************************}
\typeout{}}

\end{document}